\documentclass{jpsj-suppl}
\usepackage{txfonts} 

\newcommand{\UNEDFZERO}{\textsc{unedf0}}
\newcommand{\UNEDFONE}{\textsc{unedf1}}
\newcommand{\UNEDFTWO}{\textsc{unedf2}}
\newcommand{\UNEDF}{\textsc{unedf}}
\newcommand{\algo}{\textsc{pound}er\textsc{s}}
\newcommand{\HFBTHO}{\textsc{hfbtho}}

\newcommand{\enm}{E}
\newcommand{\rhoc}{\rho_\text{c}}
\newcommand{\knm}{K}
\newcommand{\smass}{1/M^{*}_{\rm s}}
\newcommand{\asym}{a_\text{sym}}
\newcommand{\lsym}{L}

\title{Nuclear Energy Density Optimization: UNEDF2}

\author{M.~Kortelainen$^{1,2,3,4}$, J.~McDonnell$^{3,4,5}$, W.~Nazarewicz$^{3,4,6}$,
E.~Olsen$^{3}$, P.-G.~Reinhard$^{7}$, J.~Sarich$^{8}$, N.~Schunck$^{5,3,4}$, S.~M.~Wild$^{8}$,
D.~Davesne$^{9}$, J.~Erler$^{10}$, A.~Pastore$^{11}$}
\inst{
 $^1$Department of Physics, University of Jyv\"askyl\"a, P.O. Box 35 (YFL), FI-40014 Jyv\"askyl\"a, Finland\\
 $^2$Helsinki Institute of Physics, P.O. Box 64, FI-00014 University of Helsinki, Finland\\
 $^3$Department of Physics and Astronomy, University of Tennessee, Knoxville, TN 37996, USA\\
 $^4$Physics Division, Oak Ridge National Laboratory, Oak Ridge, TN 37831, USA\\
 $^5$Physics Division, Lawrence Livermore National Laboratory, Livermore, CA 94551, USA\\
 $^6$Institute of Theoretical Physics, Warsaw University, ul. Ho\.{z}a 69, PL-00681, Warsaw, Poland\\
 $^7$Institut f\"ur Theoretische Physik, Universit\"at Erlangen, D-91054 Erlangen, Germany\\
 $^8$Mathematics and Computer Science Division, Argonne National Laboratory, Argonne, IL 60439, USA\\
 $^9$Institut de Physique Nucl\'eaire de Lyon, CNRS-IN2P3, UMR 5822, Universit\'e Lyon 1, F-69622 Villeurbanne, France\\
 $^{10}$Division of Biophysics of Macromolecules, German Cancer Research Center (DKFZ), Im Neuenheimer Feld 580, D-69120 Heidelberg, Germany\\
 $^{11}$Institut d'Astronomie et d'Astrophysique, Universit\'{e} Libre de Bruxelles - CP226, 1050 Brussels, Belgium
}

\email{markus.kortelainen@jyu.fi}

\abst{The parameters of the {\UNEDFTWO} nuclear energy density functional (EDF) model were obtained in an
optimization to experimental data consisting of nuclear binding energies, proton radii, odd-even mass staggering data, 
fission-isomer excitation energies, and single particle energies. In addition to parameter optimization,
sensitivity analysis was done to obtain parameter uncertainties and correlations.
The resulting {\UNEDFTWO} is an all-around EDF. However, the sensitivity analysis also demonstrated
that the limits of current Skyrme-like EDFs have been reached and that novel approaches are called for.}

\kword{nuclear density functional theory, Skyrme energy density}

\begin{document}
\maketitle

\section{Introduction}

The development of a universal nuclear energy density functional (EDF) capable of explaining and predicting static and 
dynamic properties of atomic nuclei is one of the important goals in the low-energy nuclear physics. This was also
one of the main research efforts in the UNEDF SciDAC-2 project \cite{(Bog13)}. During this project, the Skyrme-like EDFs
{\UNEDFZERO}~\cite{(Kor10)}, {\UNEDFONE}~\cite{(Kor12)}, and {\UNEDFTWO}~\cite{(Kor14)} were developed.
The nuclear EDF is a key element in the nuclear density functional theory (DFT).  At present, DFT is the only
microscopic theory which can be applied throughout the entire nuclear landscape.
Because parameters of the nuclear EDF cannot be precalculated with sufficient accuracy from any theory, they
must be calibrated to experimental input. 
An important aspect of the UNEDF project and the calibration of these EDFs was the joint collaboration of physicists,
applied mathematicians, and computer scientists working together toward a common goal.

With the {\UNEDFZERO} EDF we established our EDF parameter optimization procedure. By incorporating recent developments in 
optimization techniques and increased computational power, the optimization could be carried out for the first time at the 
deformed Hartree-Fock-Bogoliubov (HFB) level. Since deformation properties of {\UNEDFZERO} were found to be inadequate, 
the {\UNEDFONE} optimization paid attention to the fission properties in the actinide region. With the inclusion of data 
points on fission isomer excitation energies, the resulting {\UNEDFONE} EDF reproduced fission barriers in actinides 
well. The optimization of {\UNEDFTWO} focused on the shell structure. Here, the tensor part of the EDF was also included 
in the set of optimized parameters. To constrain tensor coupling constants, data from single-particle levels was included
in the experimental data set. In addition to parameter optimization, all {\UNEDF} parameterizations also
provided results from the sensitivity analysis.


\section{Theoretical framework}
In the Skyrme-EDF framework, the total energy of a nucleus is a functional of the one-body density matrix
$\rho$ and the pairing density matrix $\tilde{\rho}$. The total energy is a sum of kinetic energy, 
Skyrme energy, pairing energy, and Coulomb energy. The time-even part of the Skyrme energy density reads
\begin{equation} \label{eq:skyrme}
{\mathcal E}^{\rm Sk}_{t}({\bf r}) = C^{\rho}_{t}[\rho_{0}({\bf r})] \rho_{t}^{2}({\bf r}) + C^{\tau}_{t} \rho_{t}({\bf r})\tau_{t}({\bf r})
+C^{\Delta\rho}_{t} \rho_{t}({\bf r}) \Delta \rho_{t}({\bf r}) + C^{\nabla J}_{t}({\bf r}) \rho_{t}({\bf r}) \nabla \cdot {\bf J}_{t}({\bf r})
+C^{J}_{t} {\mathbb J}^{2}_{t}({\bf r}) \, ,
\end{equation}
which is composed of isoscalar ($t=0$) and isovector ($t=1$) densities. The density dependent coupling constant in 
Eq. (\ref{eq:skyrme}) is defined by
\begin{equation}
C^{\rho}_{t}[\rho_{0}({\bf r})] = C^{\rho}_{t0} + C^{\rho}_{t{\rm D}}\rho_{0}^{\gamma}({\bf r}) \, .
\end{equation}
Standard definitions of densities appearing in Eq.~(\ref{eq:skyrme}) can be found in Ref.~\cite{(Ben03)}.
The volume coupling constants, $\left\{ C^{\rho}_{t0},C^{\rho}_{t{\rm D}}, C^{\tau}_{t}, \gamma\right\}$,
can be related to the infinite nuclear matter (INM) parameters~\cite{(Kor10)}.
In all {\UNEDF} energy density optimizations the volume part was expressed by these INM parameters.
In addition to the Skyrme energy density part, the pairing term was taken to be the mixed type pairing force of Ref.~\cite{(Dob02a)}, 
with $V^{\rm n}_{0}$ and $V^{\rm p}_{0}$ as the corresponding neutron and proton pairing strengths, respectively.

The optimization of {\UNEDFTWO} was done by minimizing an objective function $\chi^{2}({\bf x})$ with respect to the
model parameters ${\bf x}$,
\begin{equation} \label{eq:chi2}
\chi^{2}({\bf x}) = \frac{1}{n_{\rm d}-n_{\rm x}} \sum_{i=1}^{n_{\rm d}} \left( \frac{s_{i}({\bf x})-d_{i}}{w_{i}} \right)^{2} \, ,
\end{equation}
where $n_{\rm d}$ and $n_{\rm x}$ are the number of data points and number of model parameters, respectively.
Furthermore, $s_{i}({\bf x})$ is the value of the $i$th observable, as predicted by the model and $d_{i}$ is the corresponding
experimental value. The experimental data set consisted of binding energies of 29 spherical and 47 deformed nuclei,
13 odd-even mass staggering (OEM) data points, 28 proton radii, 4 fission isomer excitation energies, and 
9 single-particle (sp) level energy splittings. The used experimental proton radii values were deduced from the measured charge radii.
Lastly, $w_{i}$ in Eq.~(\ref{eq:chi2}) is the weight of the $i$th observable. Here, the selected weights were $2\,{\rm MeV}$ 
for binding energies, $0.02\,{\rm fm}$ for proton radii, $0.1\,{\rm MeV}$ for OEM data, $0.5\,{\rm MeV}$ for fission 
isomer excitation energies, and $1.2\,{\rm MeV}$ for sp-level energies.


\section{The {\UNEDFTWO} energy density}

The optimization of all {\UNEDF} EDFs was carried out at the axially deformed HFB level, with the computer code 
{\HFBTHO}~\cite{(Sto13)}. This code solves the HFB equations in an axially symmetric deformed harmonic 
oscillator (HO)  basis. All the nuclei were computed in a space of 20 major HO shells.
Similarly to the {\UNEDFZERO} and {\UNEDFONE} parameterizations, the optimization was carried out by 
using the {\algo} algorithm \cite{(Wil14)}, which was found to be significantly faster compared to the traditionally used Nelder-Mead 
algorithm \cite{(Kor10)}. The {\UNEDFTWO} optimization utilized a hybrid parallel OpenMP+MPI scheme.
Similarly to the {\UNEDFZERO} and {\UNEDFONE} optimizations, the parameters of the functional were not allowed to 
attain unphysical values, so bounds were imposed on the range of variation for each parameter.

In addition to the optimization, we did a complete sensitivity analysis for the optimized 
parameter set in order to obtain standard deviations and correlations of the model parameters. 
The sensitivity analysis provides useful information about which of the parameters are strongly correlated 
and which of the parameters are poorly determined by the employed data set.
This analysis also can be used to estimate the impact of one data point on the position of the
$\chi^{2}({\bf x})$ minimum.
Most importantly, sensitivity analysis is an important tool when addressing the predictive power 
of the model and associated model uncertainties. Once the covariance matrix is known, the
model errors can be propagated \cite{(Gao13),(Kor13),(Dob14),(Sch14),(Erl14),(Kor14b)}. In particular, when an EDF model
is used in extrapolation to an experimentally unknown region, the role of the model errors becomes
prominent.

A numerical criterion, based on linear response theory in symmetric nuclear matter, was established in Ref.~\cite{(Hel13)}
to determine the eventual presence of finite-size instabilities in calculations of the nuclei. 
We verified that {\UNEDFTWO} respects this criterion, thus making it a reliable EDF for the calculation of finite nuclei.

\begin{table}[tbh]
\caption{The {\UNEDFTWO} parameterization. Listed are parameter name, parameter value, and standard 
deviation $\sigma$ for each parameter. Energy per particle ($\enm$), nuclear matter incompressibility ($\knm$), 
symmetry energy ($\asym$), and the slope of symmetry energy ($\lsym$) are in units of MeV, saturation
density ($\rhoc$) is in units of fm$^{-3}$, scalar effective mass ($\smass$) is unitless, $C^{\Delta\rho}_{t}$, 
$C^{\nabla J}_{t}$, and $C^{J}_{t}$ are in units of MeV\,fm$^{5}$, and $V_{0}^{\rm n/p}$ is in units of MeV\,fm$^{3}$.}
\label{t:unedf2}
\begin{center}
\begin{tabular}{lll|lll}
\hline
Parameter & Value    & $\sigma$ & Parameter & Value & $\sigma$ \\
\hline
$\enm$    & $-$15.8  & N/a     & $C^{\Delta\rho}_{0}$  & $-$46.831   & 2.689  \\[2pt]
$\rhoc$   & 0.15631  & 0.00112 & $C^{\Delta\rho}_{1}$  & $-$113.164  & 24.322 \\[2pt]
$\knm$    & 239.930  & 10.119  & $V^{\rm n}_{0}$       & $-$208.889  & 8.353  \\[2pt]
$\smass$  & 1.074    & 0.052   & $V^{\rm p}_{0}$       & $-$230.330  & 6.792  \\[2pt]
$\asym$   & 29.131   & 0.321   & $C^{\nabla J}_{0}$    & $-$64.309   & 5.841  \\[2pt]
$\lsym$   & 40.0     & N/a     & $C^{\nabla J}_{1}$    & $-$38.650   & 15.479 \\[2pt]
          &          &         & $C^{J}_{0}$           & $-$54.433   & 16.481 \\[2pt]
          &          &         & $C^{J}_{1}$           & $-$65.903   & 17.798 \\[2pt]
\hline
\end{tabular}
\end{center}
\end{table}

Table~\ref{t:unedf2} lists the {\UNEDFTWO} parameterization, along with the corresponding
parameter standard deviations. The volume part of the EDF, expressed by INM parameters, is listed in 
the left column. 
In the sensitivity analysis, the parameters that hit the set boundaries during the optimization were excluded.
Also, since the data set was incapable of constraining the vector effective mass, the SLy4 value 
of $1/M_{\rm v}^{*} \approx 1.250$ was used \cite{(Cha98)}.
(See the supplementary material of Ref.~\cite{(Kor14)} for precise presentation of the parameterization.)
When comparing {\UNEDFTWO} parameter uncertainties to those of {\UNEDFONE}, the {\UNEDFTWO} parameters usually have the same 
or smaller magnitude. Furthermore, the parameter uncertainty interval with {\UNEDFONE} is usually narrower 
compared to {\UNEDFZERO}, which indicates that the {\UNEDFTWO} EDF is better constrained with the current data, 
and -- given the current observables -- any major further improvement is unlikely.

\begin{figure}[htb]
\begin{center}
\includegraphics[width=0.5\linewidth]{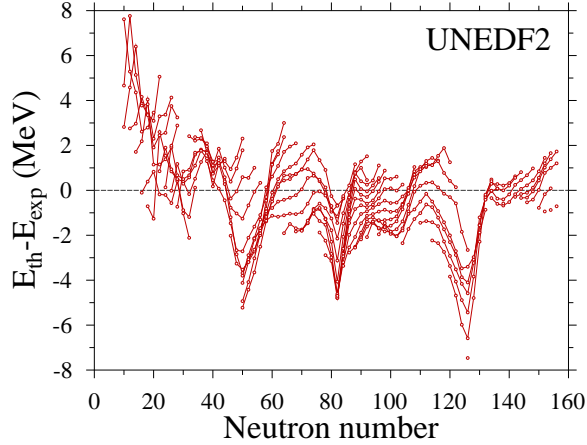}
\caption{The residuals of the even-even nuclear binding energies, calculated with {\UNEDFTWO}, compared to experimental
data of Ref.~\cite{(Aud03)}. The lines indicate isotopic chains.}\label{f:be}
\end{center}
\end{figure}

The nuclear ground-state properties predicted by the {\UNEDFTWO} EDF were computed for the whole even-even
nuclear landscape. This was done with a parallel calculation scheme where each nucleus was distributed
to a separate CPU core~\cite{(Erl12b)}. 
The same setup was also used for evaluation of the systematic error for the boundaries of the nuclear landscape,
as predicted by various Skyrme-EDF models~\cite{(Erl12)}.
Fig.~\ref{f:be} shows the calculated residuals of the even-even nuclear binding energies with the {\UNEDFTWO} EDF.
As shown, the residuals are not randomly distributed, and clear arc-like features can be seen, common to many
mean-field methods. This is one indication that Skyrme-like models are lacking some physics.
The total root-mean-square (r.m.s.) deviation from the experimental data was $1.95\,{\rm MeV}$, which is
a bit higher compared to $1.43\,{\rm MeV}$ for {\UNEDFZERO}, but similar to the $1.91\,{\rm MeV}$ for
{\UNEDFONE}. For two-neutron separation energies, the r.m.s. deviation of {\UNEDFTWO} was $0.84\,{\rm MeV}$
and for two-proton separation energies the r.m.s. deviation was $0.78\,{\rm MeV}$. 
With these observables, the r.m.s. deviation for {\UNEDFZERO} and {\UNEDFONE} was similar in magnitude.
With the proton radii, the {\UNEDFTWO} r.m.s. deviation was $0.018\,{\rm fm}$, which is about the same 
as {\UNEDFZERO} and {\UNEDFONE}.

\begin{figure}[htb]
\begin{center}
\includegraphics[width=0.85\linewidth]{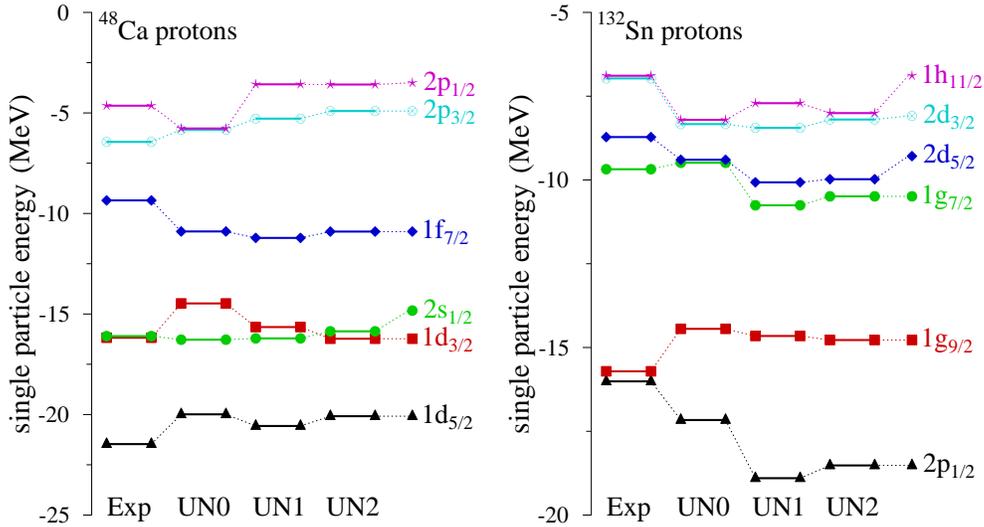}
\caption{Calculated proton single particle energies in $^{48}$Ca and $^{132}$Sn with {\UNEDFZERO} (UN0), {\UNEDFONE} (UN1), 
and {\UNEDFTWO} (UN2) EDFs compared to the experimental data of Ref.~\cite{(Sch07)}. The experimental data were
deduced from the spectra of neighboring odd-even nucleus and the binding energy differences between doubly magic nuclei
and their corresponding odd-even neighbors.}\label{f:sp}
\end{center}
\end{figure}

As mentioned, one focal point of the {\UNEDFTWO} study was the shell structure. Fig.~\ref{f:sp} shows proton 
single particle (sp) levels in $^{48}$Ca and $^{132}$Sn as calculated with {\UNEDF} EDFs. All the sp-levels here, as well as during the
optimization of {\UNEDFTWO}, were calculated within the equal filling quasiparticle blocking procedure. 
The total r.m.s. deviation of the sp-levels in doubly-magic nuclei with {\UNEDFTWO}, from the experimental levels of 
Ref.~\cite{(Sch07)}, was 1.38\,MeV. This is close to the best possible r.m.s. deviation that can be 
attained with a Skyrme-like EDF \cite{(Kor08)}. The two-particle separation energies across the shell gaps
of doubly-magic nuclei were reproduced better with {\UNEDFTWO} when compared to {\UNEDFZERO} and {\UNEDFONE}.


\section{Conclusions}

The {\UNEDFTWO} EDF was optimized to the experimental data set containing a rather large variety of 
observable types. The optimization also included tensor coupling constants, which could be constrained
due to the expanded data set used. The performance of {\UNEDFTWO} was tested against
various experimental data \cite{(Kor14)}. Global properties were found to be on par with the
previous {\UNEDFONE} parameterization. Fission properties of {\UNEDFTWO} were slightly degraded from
those of {\UNEDFONE}, particularly for the outer barrier heights. The {\UNEDFTWO} EDF can be viewed 
as a balanced all-around EDF.

In addition to parameter optimization, sensitivity analysis was done to obtain parameter uncertainties and correlations.
This sensitivity analysis clearly demonstrated that the limits of current Skyrme-like EDFs have been reached and that novel 
approaches are called for. Similar conclusions were obtained in other studies, that is,
any further major improvements with Skyrme-like EDFs are unlikely \cite{(Kor08),(Tar14)}.
To improve the current situation, new theoretical efforts have been launched. For example, the novel EDFs with higher order 
terms \cite{(Car08),(Dob12),(Rai14),(Bec14)} or enriched  density dependence \cite{(Geb10),(Sto10)} could capture more
physics and reduce systematic errors in theory.

\section*{Acknowledgments}
This work was supported by the U.S. Department of Energy 
under Contract Nos. DE-SC0008499, DE-FG02-96ER40963, and DE-FG52-09NA29461
(University of Tennessee), DE-AC02-06CH11357 (Argonne National Laboratory),
and DE-AC52-07NA27344 (Lawrence Livermore National Laboratory); by the Academy
of Finland under the Centre of Excellence Programme 2012-2017 (Nuclear and
Accelerator Based Physics Programme at JYFL) and the FIDIPRO programme; and by the
European Union's Seventh Framework Programme ENSAR (THEXO) under Grant No.
262010.

Computational resources were provided through an INCITE award ``Computational
Nuclear Structure'' by the National Center for Computational Sciences (NCCS)
and National Institute for Computational Sciences (NICS) at Oak Ridge National
Laboratory, through a grant by the Livermore Computing Resource Center at
Lawrence Livermore National Laboratory, and through a grant by the Laboratory
Computing Resource Center at Argonne National Laboratory.


\begin{thebibliography}{10}

\bibitem{(Bog13)}
S.~Bogner, A.~Bulgac, J.~Carlson, J.~Engel, G.~Fann, R.~Furnstahl, S.~Gandolfi,
  G.~Hagen, M.~Horoi, C.~Johnson, M.~Kortelainen, E.~Lusk, P.~Maris, H.~Nam,
  P.~Navratil, W.~Nazarewicz, E.~Ng, G.~Nobre, E.~Ormand, T.~Papenbrock,
  J.~Pei, S.~Pieper, S.~Quaglioni, K.~Roche, J.~Sarich, N.~Schunck,
  M.~Sosonkina, J.~Terasaki, I.~Thompson, J.~Vary, and S.~Wild: Comput. Phys.
  Comm. {\bfseries 184} (2013) 2235.

\bibitem{(Kor10)}
M.~Kortelainen, T.~Lesinski, J.~Mor\'e, W.~Nazarewicz, J.~Sarich, N.~Schunck,
  M.~V. Stoitsov, and S.~Wild: Phys. Rev. C {\bfseries 82} (2010) 024313.

\bibitem{(Kor12)}
M.~Kortelainen, J.~McDonnell, W.~Nazarewicz, P.-G. Reinhard, J.~Sarich,
  N.~Schunck, M.~V. Stoitsov, and S.~M. Wild: Phys. Rev. C {\bfseries 85}
  (2012) 024304.

\bibitem{(Kor14)}
M.~Kortelainen, J.~McDonnell, W.~Nazarewicz, E.~Olsen, P.-G. Reinhard,
  J.~Sarich, N.~Schunck, S.~M. Wild, D.~Davesne, J.~Erler, and A.~Pastore:
  Phys. Rev. C {\bfseries 89} (2014) 054314.

\bibitem{(Ben03)}
M.~Bender, P.-H. Heenen, and P.-G. Reinhard: Rev. Mod. Phys. {\bfseries 75}
  (2003) 121.

\bibitem{(Dob02a)}
J.~Dobaczewski, W.~Nazarewicz, and M.~V. Stoitsov: Eur. Phys. J. A {\bfseries
  15} (2002) 21.

\bibitem{(Sto13)}
M.~V. Stoitsov, N.~Schunck, M.~Kortelainen, N.~Michel, H.~Nam, E.~Olsen,
  J.~Sarich, and S.~Wild: Comput. Phys. Comm. {\bfseries 184} (2013) 1592.

\bibitem{(Wil14)}
S.~M. {Wild}, J.~{Sarich}, and N.~{Schunck}: arXiv:1406.5464  (2014).

\bibitem{(Gao13)}
Y.~Gao, J.~Dobaczewski, M.~Kortelainen, J.~Toivanen, and D.~Tarpanov: Phys.
  Rev. C {\bfseries 87} (2013) 034324.

\bibitem{(Kor13)}
M.~Kortelainen, J.~Erler, W.~Nazarewicz, N.~Birge, Y.~Gao, and E.~Olsen: Phys.
  Rev. C {\bfseries 88} (2013) 031305.

\bibitem{(Dob14)}
J.~Dobaczewski, W.~Nazarewicz, and P.-G. Reinhard: J. Phys. G {\bfseries 41}
  (2014) 074001.

\bibitem{(Sch14)}
N.~{Schunck}, J.~D. {McDonnell}, J.~{Sarich}, S.~M. {Wild}, and D.~{Higdon}:
  arXiv:1406.4383  (2014).

\bibitem{(Erl14)}
J.~{Erler} and P.-G. {Reinhard}: arXiv:1408.0208  (2014).

\bibitem{(Kor14b)}
M.~{Kortelainen}: arXiv:1409.1413  (2014).

\bibitem{(Hel13)}
V.~Hellemans, A.~Pastore, T.~Duguet, K.~Bennaceur, D.~Davesne, J.~Meyer,
  M.~Bender, and P.-H. Heenen: Phys. Rev. C {\bfseries 88} (2013) 064323.

\bibitem{(Cha98)}
E.~Chabanat, P.~Bonche, P.~Haensel, J.~Meyer, and R.~Schaeffer: Nucl. Phys. A
  {\bfseries 635} (1998) 231.

\bibitem{(Aud03)}
{G. Audi, A.H. Wapstra, and C. Thibault, Nucl. Phys. A {\bf 729}, 337 (2003).}

\bibitem{(Erl12b)}
J.~Erler, N.~Birge, M.~Kortelainen, W.~Nazarewicz, E.~Olsen, A.~Perhac, and
  M.~Stoitsov: J. Phys. Conf. Ser. {\bfseries 402} (2012) 012030.

\bibitem{(Erl12)}
J.~Erler, N.~Birge, M.~Kortelainen, W.~Nazarewicz, E.~Olsen, A.~Perhac, and
  M.~Stoitsov: Nature {\bfseries 486} (2012) 509.

\bibitem{(Sch07)}
N.~Schwierz, I.~Wiedenhover, and A.~Volya: arXiv:0709.3525  (2007).

\bibitem{(Kor08)}
M.~Kortelainen, J.~Dobaczewski, K.~Mizuyama, and J.~Toivanen: Phys. Rev. C
  {\bfseries 77} (2008) 064307.

\bibitem{(Tar14)}
{D. Tarpanov, J. Dobaczewski, J. Toivanen, and B.G. Carlsson, arXiv:1405.4823
  (2014)}.

\bibitem{(Car08)}
B.~G. Carlsson, J.~Dobaczewski, and M.~Kortelainen: Phys. Rev. C {\bfseries 78}
  (2008) 044326.

\bibitem{(Dob12)}
J.~Dobaczewski, K.~Bennaceur, and F.~Raimondi: J. Phys. G {\bfseries 39} (2012)
  125103.

\bibitem{(Rai14)}
F.~Raimondi, K.~Bennaceur, and J.~Dobaczewski: J. Phys. G {\bfseries 41} (2014)
  055112.

\bibitem{(Bec14)}
P.~Becker, D.~Davesne, J.~Meyer, A.~Pastore, and J.~Navarro: arXiv:1406.0340
  (2014).

\bibitem{(Geb10)}
B.~Gebremariam, T.~Duguet, and S.~K. Bogner: Phys. Rev. C {\bfseries 82} (2010)
  014305.

\bibitem{(Sto10)}
M.~Stoitsov, M.~Kortelainen, S.~K. Bogner, T.~Duguet, R.~J. Furnstahl,
  B.~Gebremariam, and N.~Schunck: Phys. Rev. C {\bfseries 82} (2010) 054307.

\end{thebibliography}

\end{document}